\documentclass[english,preprint, notitlepage, superscriptaddress,amsmath,floatfix]{revtex4-2}

\usepackage[T1]{fontenc}

\usepackage[utf8]{inputenc}

\usepackage{color}
\usepackage{units}
\usepackage{amssymb}
\usepackage{graphicx}

\usepackage{natbib}

\usepackage{babel}

\begin{document}

\title{A gate-tunable graphene Josephson parametric amplifier}

\author{Guilliam Butseraen}
\author{Arpit Ranadive}
\author{Nicolas Aparicio}
\author{Kazi Rafsanjani Amin}
\author{Abhishek Juyal}
\author{Martina Esposito}
\affiliation{Univ. Grenoble Alpes, CNRS, Grenoble INP, Institut N\'eel, 38000 Grenoble, France}
\author{Kenji Watanabe}
\affiliation{Research Center for Functional Materials, 
National Institute for Materials Science, 1-1 Namiki, Tsukuba 305-0044, Japan}
\author{Takashi Taniguchi}
\affiliation{International Center for Materials Nanoarchitectonics, 
National Institute for Materials Science,  1-1 Namiki, Tsukuba 305-0044, Japan}
\author{Nicolas Roch}
\affiliation{Univ. Grenoble Alpes, CNRS, Grenoble INP, Institut N\'eel, 38000 Grenoble, France}
\author{François Lefloch}
\affiliation{Univ. Grenoble Alpes, CEA,  Grenoble INP, IRIG, 38000 Grenoble, France}
\author{Julien Renard}\email{julien.renard@neel.cnrs.fr}
\affiliation{Univ. Grenoble Alpes, CNRS, Grenoble INP, Institut N\'eel, 38000 Grenoble, France}

\maketitle


\textbf{With a large portfolio of elemental quantum components, superconducting quantum circuits have contributed to dramatic advances in microwave quantum optics \cite{Schoelkopf2008}. Of these elements, quantum-limited parametric amplifiers \cite{Castellanos-Beltran2007,Castellanos-Beltran2008,Vijay2011}  have proven to be essential for low noise readout of quantum systems whose energy range is intrinsically low (tens of $\mu eV$). They are also used to generate non classical states of light that can be a resource  for quantum enhanced detection \cite{Backes2021}. Superconducting parametric amplifiers, like quantum bits, typically utilize a Josephson junction as a source of magnetically tunable and dissipation-free nonlinearity. In recent years, efforts have been made to introduce  semiconductor weak links as electrically tunable nonlinear elements, with demonstrations of microwave resonators and quantum bits using semiconductor nanowires \cite{DeLange2015a,Larsen2015a}, a two dimensional electron gas \cite{Casparis2018}, carbon nanotubes \cite{Mergenthaler2021} and graphene \cite{Schmidt2018,Wang2019}. However, given the challenge of balancing nonlinearity, dissipation, participation, and energy scale, parametric amplifiers have not yet been implemented with a semiconductor weak link. Here we demonstrate a parametric amplifier leveraging a graphene Josephson junction and show that its working frequency is widely tunable with a gate voltage. We report gain exceeding 20~dB and noise performance close to the standard quantum limit. Our results complete the toolset for electrically tunable superconducting quantum circuits and offer new opportunities for the development of quantum technologies such as quantum computing, quantum sensing and fundamental science \cite{Sikivie2021}.
}


In superconducting quantum circuits, information about quantum systems, for instance quantum bits (qubit) or optomechanical devices, is carried by very low power microwave signals. For efficient readout, it is necessary to amplify such signal with a minimal added noise \cite{Teufel2009,Walter2017}. Parametric amplifiers can realize such task by using nonlinearity to perform wave mixing. Several sources of nonlinearity have been used in this context, such as the kinetic inductance of disordered superconductors \cite{HoEom2012} but the most common one is the Josephson junction \cite{Zimmer1967}. Such element indeed provides a lossless nonlinearity mandatory for a low noise amplification process. Major advances in the field have been made in the last 15 years \cite{Castellanos-Beltran2007,Yamamoto2008,Bergeal2010,Macklin2015,Lecocq2017,Frattini2018,Planat2019}, allowing complementary developments of superconducting quantum bits. In the usual implementation of Josephson parametric amplifiers (JPA), the Josephson junction is coupled to a resonator to enhance its interaction with the microwave field. In such design, amplification can be achieved in a limited bandwidth around a frequency set by the resonator characteristics. This frequency can be tuned by changing the Josephson energy, acting on the total inductance $L$ of the system and hence its resonance frequency $f=\frac{1}{2\pi\sqrt{LC}}$, where $C$ is the total capacitance. Usual tunnel Josephson  junctions are made of aluminum and an aluminum oxide barrier. Superconducting quantum interference device (SQUID) geometry allows an applied magnetic field to modulate the Josephson energy \cite{Castellanos-Beltran2007}. This is also what is typically done to change the frequency of a qubit \cite{Vion2002,Koch2007,Manucharyan2009}. Recently, hybrid semiconductor/superconductor platforms have emerged with the goal of building voltage tunable qubits. In this case, the Josephson junctions are made using a semiconductor weak link, and changing the charge carrier density with a gate voltage modifies the Josephson energy, hence the qubit frequency \cite{DeLange2015a,Larsen2015a,Casparis2018,Wang2019}. Such a voltage tunable Josephson junction can in principle be used to build a voltage tunable JPA but such realization has been elusive so far. 
 In this work we demonstrate such amplifier based on a graphene Josephson junction.

\begin {figure*}[!t]
\begin{center}
    \includegraphics[width=0.9\textwidth, keepaspectratio]{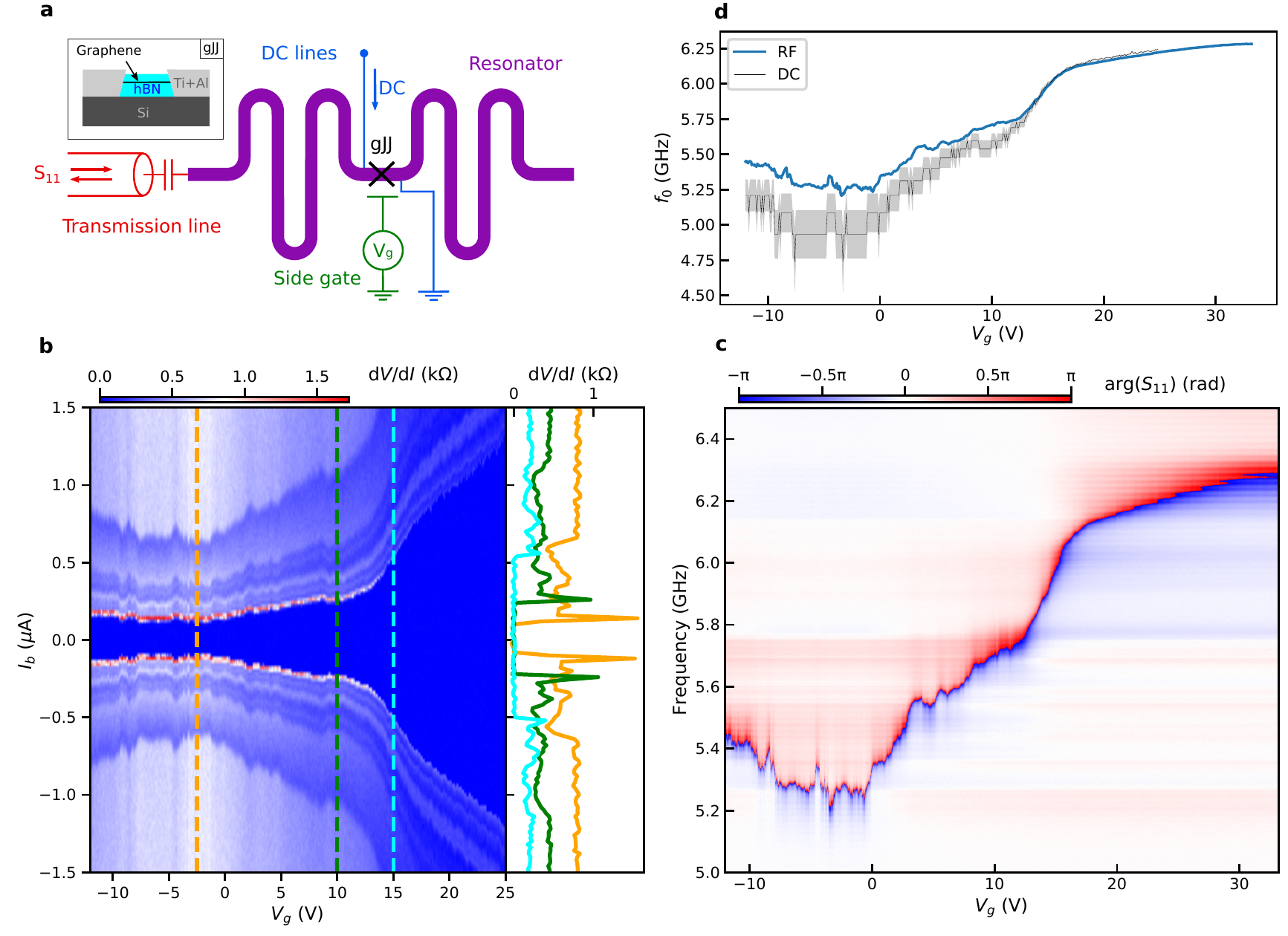} 
    \caption[]{Graphene Josephson junction embedded in a microstrip superconducting microwave resonator.
      (a): Device geometry. The system is probed in reflection (S$_{11}$ scattering parameter) via a radiofrequency (RF) transmission line connected to the resonator with a coupling capacitor. The graphene Josephson junction (gJJ) is located in the center of the resonator. Additional lines allow for DC measurements. A voltage $V_g$ on the side gate tunes the carrier concentration in the gJJ. In the inset, a cross-section of the gJJ illustrates the side contacts with Ti/Al on the h-BN encapsulated graphene.  
      (b): Differential resistance (color scale) of the gJJ as a function of $V_g$ and the bias current $I_b$. The blue region represents the dissipation-less regime, i.e. below the junction critical current. The critical current depends on the carrier concentration, controlled by $V_g$, and is smaller near the Dirac point at $V_g \approx -3~V$. Linecuts at different gate voltages (right) are used to determine the evolution of the critical current with gate voltage.  (c): Phase of the S$_{11}$ parameter as a function of gate voltage and microwave frequency. The resonance appears as a 2$\pi$ phase shift as a function of frequency.  (d): Comparison of the resonance frequency extracted from microwave reflection (c) (blue solid lines) and the one predicted from the change of Josephson inductance inferred from the critical current (b) (grey line). The error bands stem from the current bias digitization steps that lead to uncertainty in the determination of the critical current (see supplementary information, section IV).
\label{fig:1}
} 
\end{center}
\end {figure*}

We begin with a description of the device and microwave characterizations in the linear regime. In Fig.~\ref{fig:1}a, we present the schematic of the designed parametric amplifier. We use a van der Waals heterostructure made of graphene encapsulated in hexagonal boron nitride (h-BN), connected to two superconducting leads (Ti/Al) to build the Josephson junction. h-BN encapsulation and side contacts ensure high charge carrier mobility and low contact resistance which are both needed for reaching large critical currents. The junction is embedded at the voltage node of a half-wave microwave resonator with resonant frequency $\frac{\omega_0}{2\pi}$=6.44~GHz ( measured experimentally in the absence of the junction, see supplementary information, section II) and provides the nonlinearity. 
The resonator is probed at microwave frequencies in a reflection geometry. The scattering parameter S$_{11}$ is measured while additional probes allow us to characterize the junction low-frequency (DC) properties. In addition, a side gate allows to control the carrier density in the graphene junction. Measurements are made at the base temperature of a dilution refrigerator ($\sim$ 25~mK). In Fig.~\ref{fig:1}b we present the DC resistance of the device as a function of the gate voltage ($V_g$) and bias current ($I_b$). 
At low DC bias current, the device resistance vanishes and a Josephson supercurrent can flow in the graphene. Above the critical current $I_C$, dissipation kicks in, and as a result, we observe a non-zero differential resistance. We observe that the critical current strongly depends on the gate voltage \cite{Heersche2007,Calado2015} and can be varied from 100~nA  up to more than 1.3~$\mu$A. Such large critical current and the value of the $R_n\times I_C$ product ($R_n\times I_C \sim 1.4 \Delta$, where $R_n$ and $\Delta$ are the normal state resistance and the induced superconducting gap, see supplementary information, section IV)  proves the high quality of the junction \cite{Park2018}. We also observe that the gate voltage modifies the microwave resonance frequency (Fig.~\ref{fig:1}c)\cite{Schmidt2018}, that is measured with a vector network analyzer (VNA) in the low power limit. For gate voltages corresponding to a high critical current, the frequency is close to the measured bare resonance frequency $f_0=\frac{\omega_0}{2\pi}$. For lower critical currents, the frequency is reduced. This is a direct consequence of the relationship between the critical current and the Josephson inductance. Assuming for simplicity a sinusoidal current phase relation and zero phase bias across the junction, we have $L_J=\frac{\Phi_0}{2\pi I_c}$, where $\Phi_0=\frac{h}{2e}$ is the magnetic flux quantum. A modulation of $I_C$ with $V_g$ thus translates into a modification of $L_J$ resulting in a change of the resonance frequency: $\omega_r(V_g)=\frac{1}{\sqrt{(L_0+L_J(V_g))C}}$, where $L_0$ is the resonator inductance in the absence of the Josephson junction. In Fig.~\ref{fig:1}d, we compare the prediction of the resonance frequency given by this simple equivalent lumped element model with its experimental determination using microwave measurements (see supplementary information, section IV), showing a good agreement between the two. The discrepancies at low frequencies are attributed to an underestimation of the critical current in the DC experiment, which rather measures the switching current and can thus overestimate the Josephson inductance. At high frequency, on the other hand, the mismatch can be attributed to deviations of the current phase relation from its assumed sinusoidal form. In this region, our data indicate a larger inductance than the one predicted by the critical current (see supplementary information, section IV). This would point towards a forward skewed current phase relation, which has been observed in ballistic graphene Josephson junctions \cite{Nanda2017,Schmidt2020}.

The ability to tune the resonance frequency using a graphene based Josephson junction is fundamentally different than flux tuning of a SQUID \cite{Castellanos-Beltran2007}. Not only the control knob is different (i.e. electric field vs magnetic field), but the underlying mechanism for the change of inductance is also distinct. In a SQUID, the phase across the junctions depends on the magnetic flux bias, and the inductance is thus modulated by moving along the current phase relation away from zero phase. In a graphene based Josephson junction, the gate voltage modifies the critical current of the junction, hence the full current phase relation, while keeping the phase across the junction at zero. This will give more flexibility, with the possibility of an additional magnetic control of the circuit \cite{DeLange2015a} to tune independently the phase across the junction and its kinetic inductance, with interesting prospects regarding the mitigation of phase dependent dissipation processes \cite{Pop2014,Dou2021,Haller2021}.     
  
\begin {figure*}[!t]
\begin{center}
    \includegraphics[width=0.6\textwidth, keepaspectratio]{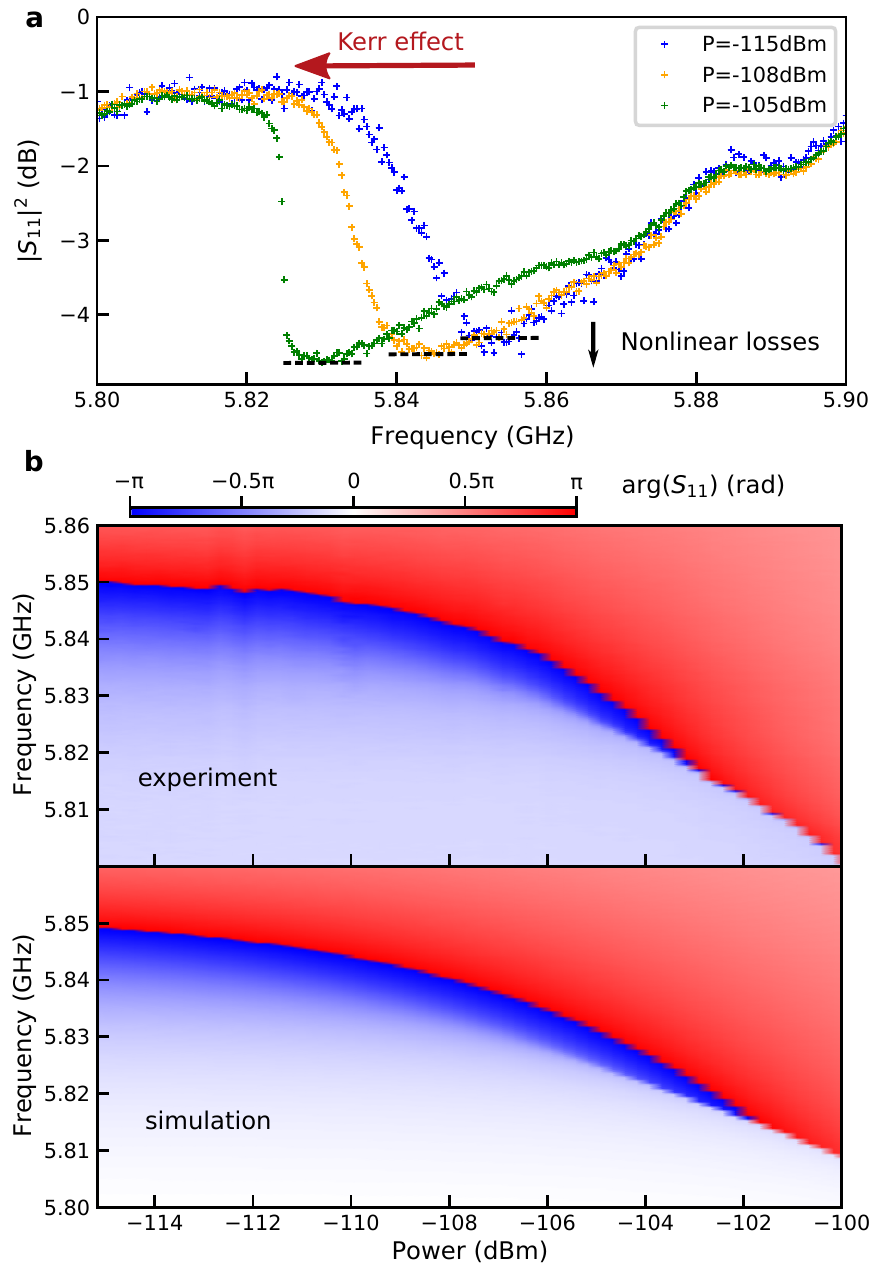} 
    \caption[]{Nonlinearity of a microwave resonator with an embedded graphene Josephson junction. Measurements of the frequency dependance of the microwave reflection magnitude (|S$_{11}$|$^2$) for different input powers (a). The resonance frequency shifts to lower frequencies (Kerr effect) and the reflection dip increases (nonlinear losses). The measurements are performed at V$_g$=15~V corresponding to a resonance frequency of $\approx$5.85~GHz in the low power regime. Measurements are made by scanning the microwave frequency from low to high, at a fixed VNA power and then changing the power. 
		(b): comparison of the measured microwave reflection (phase) with the predictions of a minimal model presented in the text. The power evolution of the resonance is captured by the simple model of Eq.~\ref{eq:1} (see text and supplementary information, section V) and gives us access to the following parameters: a low power resonance frequency $\frac{\omega_r}{2\pi}$=5.849~GHz, an external (resp. internal) energy damping rate $\gamma_1$=2$\pi\times$11.0~MHz (resp. $\gamma_2$=2$\pi\times$0.95~MHz), a Kerr coefficient K=-2$\pi\times$135~kHz and nonlinear damping rate $\gamma_3$=2$\pi\times$11~KHz.}
\label{fig:2} 
\end{center}
\end {figure*}

We now focus on the nonlinear properties of the microwave resonator with an embedded graphene Josephson junction. In Fig.~\ref{fig:2}a, we present the microwave reflection (S$_{11}$) of the device at $V_g$=15~V, with no DC bias current, for different input powers. The main feature that we observe is a decrease of the resonance frequency when the probe power increases, due to the Kerr nonlinearity. In addition, we notice that the magnitude of the reflection dip increases slightly with power. This is a signature of nonlinear losses that are not observed in usual tunnel Josephson junctions. Nonlinear losses have been shown to exist in graphene Josephson junctions and have been attributed to the presence of subgap states inside the induced superconducting gap \cite{Schmidt2020,Haller2021}, which are absent in  tunnel junctions. The position and density of these subgap states is influenced by the exact geometry of the junction \cite{Schmidt2018} and its optimization should allow to mitigate their effect in the microwave range of interest.

To model the behaviour of our device we use the following nonlinear Hamiltonian, typical for such superconducting circuits with an embedded Josephson junction \cite{Yurke2006}:
\begin{equation}
H=\hbar \omega_r A^{\dagger} A+\hbar \frac{K}{2} (A^{\dagger})^2 A^2
\label{eq:1}
\end{equation}   
where $\omega_r$ is the resonance frequency which in our case can be tuned with $V_g$, $K$ is the Kerr nonlinearity and $A$ (resp. $A^{\dagger}$)  the annihilation (resp. creation) operator of resonator photons. The resonator presents some internal losses (rate $\gamma_{2}$) and is coupled to a input/output port for measurement (coupling rate $\gamma_{1}$). We also include a nonlinear loss term, in the form of two photon losses (rate $\gamma_{3}$) \cite{Yurke2006}. 

The full nonlinear response of the system can then be calculated using input-output theory \cite{Yurke2006} (see supplementary information, section V) showing a good agreement with the experiment (see Fig.~\ref{fig:2}b). This allows us to extract the parameters of our nonlinear resonator (see Fig.~\ref{fig:2}). We note that the agreement between the model and the experiment depends on the gate voltage. For some gate voltages we see that the model is not well suited to describe the system (see supplementary information, section V). We can formulate several hypothesis to account for this. First, this could come from the fact that higher order terms of the nonlinearity are neglected in the model \cite{Kochetov2015}. Another possibility could be that the model assumes a sinusoidal current phase relation. With a non-sinusoidal current phase relation, one would need to include additional terms in the nonlinearity, in particular a cubic term that could have a contribution comparable to the quartic one. Such analysis nevertheless goes beyond the scope of the current study. Finally the presence of nonlinear losses in the device is only included to first order. The exact power dependance of the losses is likely to be more complicated and a more realistic model should be developed to take into account the detailed nature of the nonlinear losses.

The estimated Kerr nonlinearity in the device is weak, with $\frac{\lvert K \rvert}{\omega_0} \sim 10^{-4}$ , which is typical for Josephson parametric amplifiers \cite{Castellanos-Beltran2007,Bourassa2012}. Finally, above a threshold power, the resonator bifurcates and enters a bistable regime, characteristic of such a nonlinear circuit.

\begin {figure*}[!t]
\begin{center}
    \includegraphics[width=0.85\textwidth, keepaspectratio]{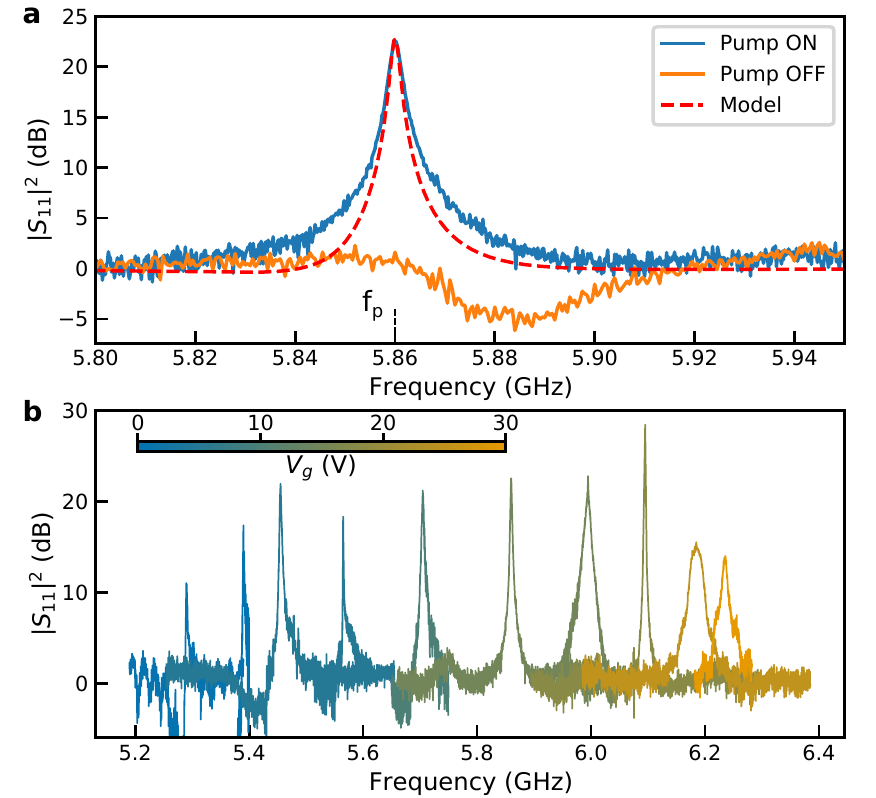} 
    \caption[]{Parametric amplification in a microwave resonator with an embedded graphene Josephson junction. 
		(a): Effect of the pump on the microwave reflection magnitude (|S$_{11}$|$^2$). The gate voltage is set at V$_g$=15~V. In absence of the pump (orange solid line, pump OFF), a dip appears at the resonance frequency. In presence of a pump at f$_\mathrm{p}$, the reflected signal shows a strong gain, up to 22~dB (blue solid line, pump ON). The measured gain is compared to the one predicted with the minimal model of parametric amplification presented in the text using the resonator parameters extracted from a one-tone measurement, a pump frequency  f$_\mathrm{p}$= 5.860~GHz and a pump power of 95\% of the critical power, corresponding experimentally to -103~dBm. A slight shift in the resonator frequency, due to an instability in the gate voltage, compared to the setpoint of Fig.~\ref{fig:2} resulted in a modified $\omega_r$=2$\pi\times$ 5.885~GHz and Kerr coefficient K=-2$\pi\times$ 111~kHz, all other parameters were kept constant. (b): Gate voltage tuning of the amplifier. The resonance frequency is tuned with $V_g$  and allows gain in a different frequency windows close to the resonance frequency by optimizing pump power and frequency for each $V_g$.
\label{fig:3} } 
\end{center}
\end {figure*}

We will now focus on the regime of interest for parametric amplification, that is slightly below bifurcation. Here, two signals are sent to the input of the device: a strong pump tone (frequency $f_p$) and a weak probe tone (frequency $f_s$), whose reflection is measured with a VNA. Both signals frequencies are chosen close to the resonance frequency of the resonator and we have the relation $2f_p=f_s+f_i$, where $f_i$ is the frequency of the complementary idler mode. We are thus in a four-wave mixing scheme of amplification which is the one expected from the Hamiltonian in Eq~\ref{eq:1}. In Fig.~\ref{fig:3}a, we see the effect of the pump on the magnitude of the reflection coefficient; in the presence of the pump, the probe signal is amplified by as much as 22~dB. This is the signature of parametric amplification. Amplification is present in a small frequency range near the pump frequency, limited by the resonator bandwidth. We extract a gain-bandwidth product of 33~MHz, which is typical for resonant Josephson parametric amplifier and mainly set by the coupling to the measurement port. In Fig.~\ref{fig:3}a, we compare the experimentally measured gain with the one predicted by the input-output model  using the parameters extracted from the one-tone measurement (see Fig~\ref{fig:2}). We see that the gain profile is qualitatively described by the formalism of Josephson parametric amplifiers \cite{Yurke2006} (see supplementary information, section V, for more details about the modeling of the gain).

In a resonant Josephson parametric amplifier, amplification can typically be achieved only in a narrow frequency range close to the resonance frequency of the superconducting resonator. In our device, the resonance frequency can be tuned with a simple gate voltage as we reported in Fig.\ref{fig:1}~. We now explore the possibility to realize amplification in a broad frequency range and present the main result of this work. In Fig.\ref{fig:3}b~, we show gain profiles measured across the full range of tunability of the resonance frequency. For each frequency, set by a chosen $V_g$, pump frequency and power are optimized accordingly. We see that we can have a large gain (>15dB) in a frequency range of about 1 GHz, i.e. more than 100$\times$ the bandwidth of the amplifier. This gives interesting perspectives for multiplexing as the amplifier could be used to selectively readout superconducting circuits at different operating frequencies depending on the gate voltage. A gate voltage tuning has advantages compared to traditional magnetic flux tuning obtained in SQUID as the local character of the electrical control will suppress crosstalk issues between different parts of the device.

Variations in the apparent gain and deviations from an ideal Lorentzian shape are attributed to non-optimal pump parameters that can come from gate voltage instability (see supplementary information, section III). Nevertheless, maximum gains are expected to be reduced when the graphene Josephson junction nonlinearity is not optimal, for instance too large, close to the Dirac point, i.e. at low frequency. We also observe that the nonlinear losses seem to increase at high frequency, above 6.2~GHz, which could explain the lower gains in this region.

\begin {figure*}[!t]
\begin{center}
    \includegraphics[width=0.85\textwidth, keepaspectratio]{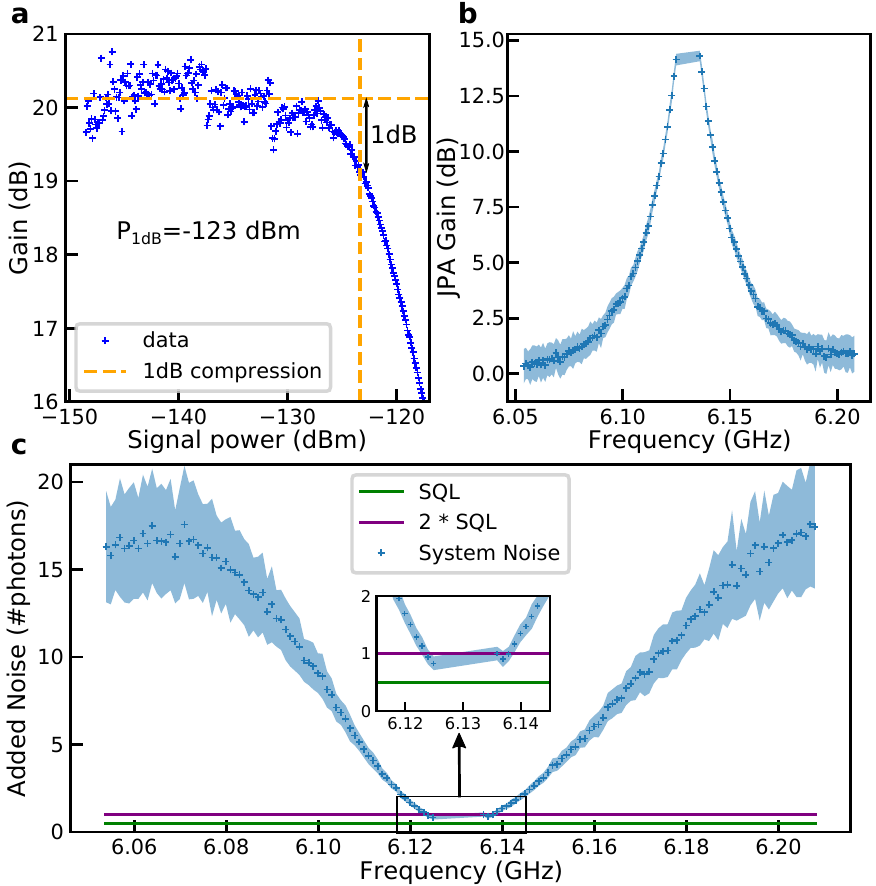} 
    \caption[]{Performance of the resonant Josephson parametric amplifier. 
		(a): Signal power dependance of the gain showing a 1~dB compression point P$_{1dB}$=-123 dBm for a gain of 20~dB at a working frequency of 6.13~GHz. 
		(b): Gain of the amplifier measured with the SNTJ and the spectrum analyzer. 
		(c): Demonstration of noise close to the standard quantum limit. The device is set such that the gain is 20~dB at 6.13~GHz (pump frequency). Away from the center of the amplification window of the device, the added noise is about 15 photons, mainly set by the HEMT amplifier at the 4~K stage. When the gain of the parametric amplifier increases, the added noise decreases and reaches values close to the standard quantum limit. Analysis is not possible at the very center of the amplification band due to the presence of the pump. The shaded area represents the error band in the determination of the added noise (see supplementary information, section VI ).
\label{fig:4} } 
\end{center}
\end {figure*}

Having demonstrated large gain and tunability we now turn to two essential characteristics of a superconducting parametric amplifier: dynamic range and noise. The dynamic range indicates the input power that can be sent to the amplifier before the gain is appreciably reduced. To measure it experimentally, we measure the gain as a function of probe power for fixed frequency and pump power. In Fig.\ref{fig:4}a~, we see that the gain is reduced by 1~dB for an input power of -123 $\pm$3~dBm: this is the 1~dB compression point P$_{1dB}$ for our device.
Such a value is comparable to the best values obtained with single Josephson junction JPA \cite{Mutus2013}. It is intrinsically related to the junction nonlinearity, critical current and the exact resonator design. This can be improved in future realizations by using arrays of junctions with larger critical currents \cite{Planat2019}.

The noise of a measurement chain designed to measure very low microwave signals is one of its key characteristics. As such, amplifiers in this chain, and especially the first one, which will often determine the noise figure of the full chain, should add as little noise as possible. For a phase preserving amplifier, the minimal amount is set by quantum mechanics which states that at least half a photon of noise, i.e. an energy of $\frac{\hbar \omega}{2}$, will be added for each additional mode coupled to the circuit \cite{Caves1982}. In a perfect Josephson parametric amplifier, working in a non degenerate mode, one can then expect a minimal added noise level of  $\frac{\hbar \omega}{2}$, since the idler mode is required in the amplification process. This is the standard quantum limit (SQL) and corresponds to a noise temperature of about 145~mK at a frequency of 6~GHz. 
 
We have seen previously that our device presents some internal losses, especially at strong pump power (nonlinear losses) which means that energy is transferred to other modes. In principle this coupling to other modes, if large enough, could degrade the noise figure of the amplifier as these modes will inject their vacuum noise into the amplifier \cite{Eichler2014}. In this case, the resulting noise is expected to be increased by a factor $\frac{\gamma_1+\gamma_{\mathrm{tot}}}{\gamma_1}$, where $\gamma_{tot}$ is the total internal loss rate. In our case, considering the internal losses of the device, we thus expect this mechanism to enhance the noise by  about 20\% above the SQL.  We measure the noise of the amplifier using a shot noise tunnel junction (SNTJ) \cite{Roy2015}, which serves as a self calibrated broadband noise source when voltage-biased  (see supplementary information, sestion I, for details about the setup) and replaces the VNA input tone. 
The broadband SNTJ output is amplified by the device and we measure the resulting signal with a spectrum analyzer. 
Studying the power spectral density of the output signal as a function of the SNTJ voltage bias, we are able to extract, for each frequency, the system gain and the noise added by the system (see supplementary information, section VI). In Fig.\ref{fig:4}c~, we present the frequency dependence of the added noise, when the amplifier is set to operate at a center frequency 6.13~GHz, together with the amplifier gain in Fig.\ref{fig:4}b~. We see that there is a clear anticorrelation between the two. When the gain of the  parametric amplifier is low, the noise added by the system is, as expected, close to 15 photons, limited by the High Electron Mobility Transistor (HEMT) amplifier noise and additional losses in the system. When the gain of the amplifier increases, getting closer to the pump frequency, the system added noise decreases dramatically and reaches a value close to the SQL. 
We can thus conclude that our graphene based Josephson parametric amplifier adds a minimal amount of noise, set by quantum mechanics, which is comparable to the state of the art of quantum limited amplifiers. 

In conclusion, we have demonstrated a Josephson parametric amplifier using an electric field tunable Josephson junction made of graphene. The amplification bandwidth of the device is tunable over about 1~GHz with a simple electric field in addition to presenting a gain > 20~dB and noise close to the standard quantum limit. This first experimental demonstration of a  gate-tunable semiconductor weak-link Josephson parametric amplifier opens interesting perspectives for different amplification schemes. For example, pumping the gate voltage at twice the signal frequency would modify the critical current and thus directly modulate the kinetic inductance. In practice such scheme should produce a three wave mixing amplification process, similar to flux pumping in SQUID based Josephson parametric amplifiers \cite{Yamamoto2008}. The three wave mixing process, which has the advantage of frequency separation between signal and pump, could also be observed in graphene based Josephson parametric amplifier using other strategies. The possibility to current bias the junction, as is demonstrated in this work, could naturally give access to the nonlinearity necessary to achieve three wave mixing \cite{Vissers2016}. In addition, in Josephson junctions presenting a non sinusoidal current phase relation, which has been reported in ballistic graphene \cite{Nanda2017,Schmidt2020}, three wave mixing terms are present even at zero DC current bias and could be directly used. Together with the recent developments regarding qubits \cite{Larsen2020}, bolometers \cite{Lee2020,Kokkoniemi2020} and the use of van der Waals materials in superconducting quantum circuits \cite{Antony2021,Antony2021b,wang2022}, the demonstration we report here for quantum limited amplifiers position semiconductor based Josephson junctions as key elements for future integrated superconducting quantum circuits.


\section*{A\lowercase{cknowledgments}}
We thank José Aumentado and Florent Lecocq (National Institute of Standards and Technology, Boulder, Colorado, USA) for providing the SNTJ and for discussions.
This work was supported by the French National Research Agency (ANR) in the framework of the Graphmon project (ANR-19-CE47-0007). K.W. and T.T. acknowledge support from the Elemental Strategy Initiative conducted by the MEXT, Japan (Grant Number JPMXP0112101001) and  JSPS KAKENHI (Grant Numbers 19H05790, 20H00354 and 21H05233). JR acknowledges E. Eyraud and W. Wernsdorfer for help with the cryogenic system. We acknowledge the work of J. Jarreau, L. Del-Rey and D. Dufeu for the design and fabrication of the sample holders and other mechanical pieces used in the cryogenic system. We thank the Nanofab group at Institut Néel for help with devices fabrication. We thank K. W. Murch and B. Sacépé for discussions and comments on the manuscript.

\section*{A\lowercase{uthor contributions}}
K. W. and T. T. grew the h-BN crystals. G. B. and J. R. designed the samples. G. B. and N. A. fabricated the devices. G. B., A. J. and J. R. performed DC measurements. G. B. performed the microwave measurements with help from K. R. A. and J. R. Noise measurements were realized by  G. B., A. R. and M. E. with help from N. R. and J. R. Data analysis was performed by G. B. with help from A. R., N. R. and J. R. The project was supervised by F. L. and J. R. G.B. prepared the figures of the manuscript. J. R. wrote the manuscript with input from all authors.

\section*{C\lowercase{ompeting Interests statement}}
N.R. is founder and share holder of Silent Waves.

\section*{M\lowercase{ethods}}
\subsection*{Devices fabrication}
The h-BN encapsulated graphene stacks were made using a polymer-free assembly technique \cite{Wang2013} on a high-resistivity Si substrate.
Devices were processed with 2 steps of e-beam lithography (80 kV) using positive resists (PMMA) and an additional layer of conductive resist for the contact step (AllResist/Electra).  Etching was performed using reactive ion etching and a mixture of O$_2$ and CHF$_3$. Metal (Ti/Al; 5~nm/60~nm) was deposited using e-beam evaporation. The superconducting contacts were made using an etch-fill technique. The size of the gJJ is 300~nm between the superconducting contacts and 1.5~$\mu$m in the transverse direction.
\subsection*{DC and Microwave measurements}
Measurements were performed in dilution refrigerators with base temperatures of ~25~mK. DC measurements were performed using low frequency lock-in techniques. The microwave measurement setup is presented in details in the Supplementary Information, section I. One tone microwave measurements were performed with a vector network analyzer. Two-tone measurements (for instance to measure the amplifier gain) were performed using an additional microwave source. The reflected signal is split with a circulator or a directionnal coupler depending on the setup and amplified at 4~K and again at room temperature. For noise measurement the signal is measured with a  spectrum analyzer. 
\subsection*{Data Availability}
The datasets generated during the current study are available from the corresponding author on reasonable request.  

\section*{R\lowercase{eferences}}


%

\end{document}